\pdfoutput=1
\documentclass[twocolumn,superscriptaddress,floatfix,prX]{revtex4-1}
\usepackage{graphics,amssymb,amsmath,epsfig,color}
\usepackage{graphicx}

\usepackage[utf8]{inputenc}
\usepackage{lineno,hyperref}
\modulolinenumbers[5]

\usepackage{times}

\usepackage[activate={true,nocompatibility},final,tracking=true,kerning=true,spacing=true,factor=1100,stretch=10,shrink=10]{microtype}

\usepackage{listings} 
\usepackage{mdframed} 
\usepackage{colortbl}
\usepackage[table]{xcolor}
\definecolor{colororca}{rgb}{1,0.93,0.8}
\definecolor{colororca2}{rgb}{0.9764, 0.9059, 0.7804}
\definecolor{colorg09}{rgb}{0.95,0.91,0.9}
\definecolor{colorg092}{rgb}{0.9764,0.9373,0.9294}
\definecolor{colornwchem}{rgb}{0.91,0.93,0.94}
\definecolor{colornwchem2}{rgb}{0.949,0.9686,0.9765}

\definecolor{light-gray}{gray}{0.87}
\usepackage[left=2.00cm, right=2.00cm,top=3cm,bottom=3cm]{geometry}

\begin{document}
\title{Revisiting the thermochemistry of chlorine fluorides}

\author{Hernán R. Sánchez}

\affiliation{Centro de Química Inorgánica,  Facultad de Ciencias Exactas, Universidad Nacional de La Plata, CONICET, CC 962, 1900 La Plata,Argentina}

\email{hernan.sanchez@quimica.unlp.edu.ar}

\begin{abstract}
In this work, accurate calculations of standard enthalpies of formation of chlorine fluorides (ClF$_n$, n=1--7; Cl$_2$F and Cl$_3$F$_2$) were performed through the isodesmic reactions scheme. It is argued that, for many chlorine fluorides, the gold standard method of quantum chemistry (CCSD(T)) is not capable to predict enthalpy values nearing chemical accuracy if atomization scheme is used. This is underpinned by a thorough analysis of total atomization energy results and the inspection of multireference features of these compounds.  Other thermodynamic quantities were also calculated at different temperatures. In order to complement the energetic description, elimination curves were studied through density functional theory as a computationally affordable alternative to highly correlated wave function-based methods. 

\end{abstract}
\maketitle

\section{Introduction}

Chlorine fluorides are very reactive species with high potential uses as oxidizers. In these compounds the chlorine atom  has a formal oxidation state ranging from +I to +VII, allowing a wide coordination numbers variety.  Currently there are limited experimental thermochemical data available about these species. On the other hand many theoretical studies have been carried out for years\cite{gropen1982electrbondclf3,pettersson1983binding,pershin1987ab,jasien1992stabilities,doi:10.1080/002689796173949,ricca2000heats,law2002thermochemistry,chen2009bonding,du2011theoretical,doi:10.1021/ic301438b,chen2013high}. Accurate prediction of thermochemical properties is crucial for the understanding of the rich chemistry of this class of compounds. Some chlorine fluorides present extreme reactivity, a solid understanding of these compounds is essential when considering the hazards and difficulties involved in their experimental manipulation.

Most of these species, if not all\cite{chen2009bonding,chen2013high}, show hypervalent character, i.e., they do not satisfy the traditional theories of Lewis and Langmuir. It is known that properties that depend of the electronic structure tend to be more difficult to calculate for hypervalent species\cite{deng2010hartree} and,  as shown in this contribution, chlorine fluorides are good examples of this fact.

Today, highly accurate molecular calculations are commonly considered almost synonymous of those performed with the so called ``gold-standard'' method in quantum chemistry, i.e., coupled cluster singles and doubles approach \cite{bartlett1981many,knowles1993coupled}  with perturbational triples contribution (CCSD(T))\cite{raghavachari1989fifth,watts1993coupled}, using very large basis sets. This is so because this method reaches energy values within chemical accuracy for most modest sized molecules (see for example \cite{martin1997benchmark}) in which the dynamical type correlation dominates. Going beyond this level of theory is neither  affordable nor necessary in most cases. A lot of cheaper composite methods to approximate its values have been developed and used widely. 

Among the nine molecules studied, experimental values are available for only three of them. This difficult the comparison between different theoretical estimations. In recent works the formation enthalpies of some of the above mentioned compounds have been calculated, using both: CCSD(T) with near complete basis set (CBS) values\cite{ricca2000heats,doi:10.1021/ic301438b}, and or composite methods\cite{jasien1992stabilities,       du2011theoretical,law2002thermochemistry}. In the early work the results for molecules with experimental values show very large errors. Even in the recent works, the error for the largest of them is still too high\cite{doi:10.1021/ic301438b,du2011theoretical,law2002thermochemistry}. This discrepancy at the CCSD(T) level of theory should  arise from an important contribution to the energy from the higher excitations not accounted in the calculations. It should be noted that, when higher excitations contribute significantly to the molecular energy, there are not good error compensation in atomization reactions if the former are neglected. Due to seemingly similar chemical environments, it can be suspected that this behaviour take place in other compounds of the series. 

The main goal of this contribution is to suggest more reliable values for the formation enthalpies of these species. To that end, unlike previous works, the isodesmic reaction scheme\cite{wheeler2012homodesmotic} is employed in pursuit of large systematic error compensation. The results obtained represents a very clear improvement for chlorine pentafluoride. In order to reinforce the idea that the improvement extend across the series, the A$_\lambda$ and \%TAE[T] diagnostics for static correlation were performed because they correlate well with higher excitations\cite{fogueri2013simple,karton2011w4-11}. As they suggested severe to pathological multireference character and, multireference effects tends to increase the importance of large excitations\cite{karton2011w4-11}, orbital occupancies of a multiconfigurational treatment were also inspected. 

The Cl$_2$F and Cl$_3$F$_2$ were added to the ClF$_n$ (n=1--7) series to include species with Cl-Cl bonds that could arise in some reaction mechanisms involving fluorine chlorides\cite{san1982kinetics}. 

The paper is structured as follows: In Section \ref{sec metodos} many methods used in this work for the calculation of enthalpies of formation are described. The Subsection \ref{sec clf2} is devoted to the accurate calculation of the small ClF$_2$ molecule, which is also useful for use in other isodesmic reactions. In Section \ref{sec results and discussion} the results are presented and it is argued that their anomalous discordance suggest the need of avoiding the atomization scheme in conjunction with the composite methods used. In Subsection \ref{sec multiconfigurational diagnostic} results of standard diagnostics of multi-configurational character are shown and an inspection of orbital occupancies is performed. The ClF$_3$ and the very difficult ClF$_7$ cases are analyzed in Subsection \ref{sec ClF3 y ClF7}. A summary of the values recommended in this contribution can be found in Subsection \ref{sec recommended}, while in Subsection \ref{sec comparison previous results} the values of this work are compared with previous results.  In Subsection \ref{sec standard thermodynamic functions} calculated values for standard thermodynamic magnitudes as functions of temperature are presented.  The difficulties in retrieving correlation energy  of fluoride chlorides give raise to uncommon failures in the description of some potential energy elimination curves, it is briefly depicted in Subsection \ref{sec potential energy curves}. Finally, in Subsection \ref{sec dft calculations} many density functionals have been tried as low cost alternatives to post-Hartree-Fock methods for non equilibrium geometries.

\section{Methods\label{sec metodos}}

The most used method for calculating standard enthalpies of formation is based on total atomization energies (TAE), and it is described, for example, in Nicolaides's work \cite{Nicolaides1996}. Within this approach, accurate values for differences of energies are necessary  for retrieving accurate enthalpy values. This method does not require further information from other molecules, and the required atomic heat of formation are well known for many chemical elements. One disadvantages is that the error compensation  which occurs in the atomization reaction is not good enough to reach chemical accuracy for some reactions, even for the gold standard method. Instead, isodesmic reaction scheme\cite{bakowies:144113}, when feasible, provides highly improved systematic error compensation, including those related to basis set extension and poor treatment of correlation energy. However, it requires the usage of accurate experimental or calculated data from molecules included in the isodesmic reaction.

The following isodesmic and isogyric reactions were used in this work for the calculation of the enthalpies of formation of the first compound to the left in each reaction

\begin{table}[htbp]
\begin{tabular}{lcl}
	ClF$_4$ + ClF&$\leftrightarrow$&ClF$_2$ + ClF$_3$\\
	ClF$_5$ + ClF &$\leftrightarrow$&2 ClF$_3$\\
	ClF$_6$ + 2 ClF&$\leftrightarrow$&ClF$_2$ + 2 ClF$_3$\\
	ClF$_7$ + 2 ClF&$\leftrightarrow$&3 ClF$_3$\\
	Cl$_2$F + ClF&$\leftrightarrow$&ClF$_2$ + Cl$_2$\\
	Cl$_3$F$_2$ + 3 ClF &$\leftrightarrow$& ClF$_2$ +2 Cl$_2$ + ClF$_3$\\
\end{tabular} 
\end{table}

\noindent
The absence of reactions for ClF, ClF$_2$ is due to their enthalpies are the most accurately known among these compounds, and will be used to make predictions about the other species. If the atomization scheme is used, the electronic structure methods employed in this work return more disperse values for ClF$_5$ than for ClF$_3$. In order to clearly illustrate the advantages of the isodesmic approach, as a first instance, the standard enthalpy of formation of the ClF$_5$ molecule was calculated by using this approach.  In Section \ref{sec ClF3 y ClF7} it is argued that a better approach is to estimate the ClF$_3$ enthalpy by means of the ClF$_5$ value.
 
The accepted values of standard enthalpies of formation, at 298.15 K, for the isodesmic procedure are: -13.308 $\pm$ 0.014 kcal$\cdot$mol$^{-1}$ for ClF\cite{atct,ruscic2005active}. For ClF$_2$ and ClF$_3$ they are -10.7 and -39.6 kcal$\cdot$mol$^{-1}$ respectively. The later values come from this work and are extensively discussed in the following sections.

It has been noted\cite{ricca2000heats} that basis set requirements increase going from ClF to ClF$_3$, when CCSD(T) is used, and that accurate atomization calculations need at least aug-cc-pV(5+d)Z\cite{dunning2001gaussian} basis sets quality combined with basis sets extrapolation schemes. Heavier species seem to be even more demanding. In this work, it was necessary to use cheaper methods, which at the same time is interesting because they include different empirical corrections, although many of these correction cancels out in the isodesmic cases.

In this work a variety of composited methods were used: (CBS-QB3\cite{montgomery1999complete,montgomery2000complete}, G3MP2B3\cite{baboul1999gaussian}, G3B3\cite{baboul1999gaussian} and G4\cite{curtiss2007gaussian}) for isodesmic reactions scheme. For comparative purposes, the density functional theory (DFT) \cite{hohenberg1964inhomogeneous,kohn1965self} was also used at this stage, with the widely used M06-2X functional\cite{zhao2008m06} and Pople's 6-311+G(3df) basis set\cite{mclean19800contracted,clark1983efficient,frisch:3265}.  

Formation enthalpy was also calculated for the above methods via total atomization scheme. Experimental atomic values at 0 K, and thermal corrections to enthalpy were taken from ref. \cite{ochterski2000thermochemistry}. Atomic spin-orbit corrections were taken from ref. \cite{curtiss2007gaussian}. Within this calculation scheme, two bond additivity correction (BAC) procedures, BAC-G3MP2B3\cite{anantharaman2005bond} and BAC-G3B3\cite{anantharaman2005bond} were included too. They lose many of their advantages in the framework of isodesmic reaction procedures because of term cancellation, which  is the reason why they were omitted before.

\subsection{The ClF$_2$ case \label{sec clf2}}

Since the ClF$_2$ radical is included in four isodesmic reactions, its formation enthalpy is particularly important here. Due to lack of experimental values, very high level of theory is needed to estimate the value for this radical. Fortunately, this species has few enough valence electrons to be able to perform high level frozen-core calculations on it. W1U\cite{barnes2009unrestricted} and W1BD\cite{barnes2009unrestricted} methods were used here and both returned formation enthalpies of -10.8 kcal$\cdot$mol$^{-1}$. Ricca\cite{ricca2000heats} obtained a value of -9.0 kcal$\cdot$mol$^{-1}$ for this property. 
 
The following unrestricted calculation procedure was carried out in this work: the optimized equilibrium geometry was found at the CCSD(T)/aug-cc-pV(Q+d)Z\cite{dunning2001gaussian} level of theory. Hartree-Fock's limit energy was approximated by using the exponential two point scheme with X dependent alpha as described for Halkier et al.\cite{halkier1999basis},  with the aug-cc-pVXZ basis sets (X = 5, 6). Also the Karton-Martin extrapolation scheme\cite{karton2006comment} was tried but negligible differences were found for the purpose. The valence correlation energy was calculated at both CCSD(T)/aug-cc-pV(Q+2df)Z and CCSD(T)/aug-cc-pV(5+2df)Z levels of theory, and a complete basis set extrapolation was done following Helgaker's et al. method\cite{helgaker1997basis,halkier1998basis}. The vibrational zero point energy was taken from the previous W1s calculations. Core correlation and relativistic effects were computed as the difference between values from CCSD(T)/aug-cc-pCVTZ-DK\cite{prascher2011gaussian} with all electrons correlated plus Douglas-Kroll-Hess 2nd order scalar relativistic method\cite{douglas1974quantum,hess1985applicability,hess1986relativistic,jansen1989revision}, and the frozen core non-relativistic one. Spin-orbit corrections were carried out for atomic energies, and values were taken from \cite{curtiss2007gaussian}. The thermal correction to standard enthalpy of ClF$_2$ was provided by the W1s calculations. This procedure gives -8.9 kcal$\cdot$mol$^{-1}$ for the standard enthalpy of formation of ClF$_2$. This value is almost equal to the one reported by Ricca. The big difference between it and those of W1s methods is noticeable being that the authors of those methods reported mean absolute error of 0.30 kcal$\cdot$mol$^{-1}$ for them. All the computational calculations above have been carried out with the Gaussian 09 software package\cite{g09}.

The effect of higher excitations was accounted as the energy difference between unrestricted valence-only calculations of CCSDT(2)$_Q$\cite{hirata2004combined} and CCSD(T), using cc-pVDZ\cite{dunning1989gaussian,woon1993gaussian} basis set. Such small basis set has been found useful for predicting energetic contribution due to high excitations\cite{bomble2006high,harding2008high}. Calculations were carried out with NWChem 6.1\cite{valiev2010nwchem}. The additive correction factor is found to be as large as -1.8~kcal$\cdot$mol$^{-1}$. Then, the recommended value for the standard formation enthalpy at 298.15 K of ClF$_2$ is -10.7~kcal$\cdot$mol$^{-1}$. Coincidence with W1s value seems to be fortuitous.

Although the Cl$_2$F has the same number of valence electrons, its basis set is slightly larger, and this prevented the CCSDT(2)$_Q$ calculations to be made.

\section{Results and discussion \label{sec results and discussion}}

The composite methods used in this work have been applied successfully in the accurate calculation of standard enthalpies of formation through atomization scheme. The mean absolute errors (MAE) in the heat of formation for G3MP2B3, G3B3 and CBS-QB3 are  1.13, 0.93 and 1.08 kcal$\cdot$mol$^{-1}$ in the G2/97 test set\cite{baboul1999gaussian,montgomery2000complete} respectively. The G4 method present a MAE of 0.80 kcal$\cdot$mol$^{-1}$ for the heats of formation included in the G3/05 test set\cite{curtiss2007gaussian}. Excluding the hydrogen containing molecules, the G3MP2B3, G3B3 and G4 values increase to 1.99, 1.65 and 1.13 kcal$\cdot$mol$^{-1}$\cite{baboul1999gaussian,curtiss2007gaussian}. The BAC-G3MP2B3 and BAC-G3B3 methods gave MAEs of 0.96 and 0.91 kcal$\cdot$mol$^{-1}$. All this values correspond to the atomization scheme.

\begin{table*}[htbp]
	\begin{center}
		\begin{tabular}{lccccccccc}
		\hline
			Method & ClF & ClF$_2$ & ClF$_3$  & ClF$_4$  & ClF$_5$  & ClF$_6$  & ClF$_7$ & Cl$_2$F & Cl$_3$F$_2$ \\ 
		\hline
			M06-2X & -12.6 & -6.0 & -29.6 & -22.1 & -44.1 & -27.14 &  33.9 & 13.4 & 24.4\\ 
			G3MP2B3& -11.6 & -8.8 & -33.2 & -24.2 & -45.5 & -29.3 & 26.5 & 13.5 & 23.9\\ 
			G3B3 & -11.9  & -9.3 & -34.9 & -26.4 & -48.7 & -34.8 &22.1 & 13.7 & 22.5\\ 
			CBS-QB3 & -13.6 & -12.0 & -36.3 & -31.1 & -49.8 & -40.7 & 20.5 & 9.6 & 16.4\\ 
			G4 & -13.4 &-12.6  & -38.6 & -31.8 & -54.6 & -41.4 & 16.2 & 11.8 & 18.9\\ 
			BAC-G3MP2B3& -12.6 & -11.2 & -38.2 & -31.7 & -56.9 & -44.1 & 8.1 & 10.3 & 16.6\\ 
			BAC-G3B3& -13.4 & -11.8 & -39.2 & -32.2 & -56.9 & -44.7 & 10.6 & 10.7 & 18.8\\ 
		\hline
		\end{tabular} 
		\caption{Standard formation enthalpies at 298.15 K, in kcal$\cdot$mol$^{-1}$, via total atomization energy scheme.\label{tabla atomizacion}}
	\end{center}
\end{table*}

\begin{table*}[htbp]
	\begin{center}
		\begin{tabular}{lcccccc}
			\hline
			Method &  ClF$_4$  & ClF$_5$  & ClF$_6$  & ClF$_7$ & Cl$_2$F & Cl$_3$F$_2$ \\
			\hline
			M06-2X 		& -32.1 & -55.3 & -44.3 & 17.4 & 9.1 & 15.1 \\ 
			G3MP2B3	 & -30.8 & -56.6 & -40.5 & 10.7 & 12.6 & 19.2 \\ 
			G3B3 		  & -31.2 & -56.8 & -43.0 & 10.6 & 11.8 & 15.4 \\ 
			CBS-QB3    & -33.3 & -56.6 & -46.4 & 10.2 & 12.4 & 17.3\\
			G4               & -31.0 & -56.6 & -41.7 & 13.0 & 11.9 & 18.3\\ 
			\hline
		\end{tabular} 
		\caption{Standard formation enthalpies at 298.15 K, in kcal$\cdot$mol$^{-1}$, via isodesmic reaction scheme.\label{table isodesmicas}}
	\end{center}
\end{table*}

They were used before for chlorine fluorides\cite{du2011theoretical,law2002thermochemistry} without good success, at least, for the last members of the series. The corresponding values were calculated and can be found in Table \ref{tabla atomizacion}. 

The fact that the mean of the absolute difference between the heats of formation, obtained with two methods for each molecule,  far exceed those expected for the employed methods, implies abnormal behavior of at least one method. This is aggravated because there is a clear correlation between the errors of different methods. For example, in the Supporting Information of ref. \cite{anantharaman2005bond}  there is a lists of heat of formation of 46 compounds for which the BAC-G2 method has deviations higher than 1 kcal$\cdot$mol$^{-1}$. For them, the MAE of G3MP2B3 and BAC-G3B3 is 2.4 and 1.9 kcal$\cdot$mol$^{-1}$, while the MAE of the difference of errors of the methods is 1.2 kcal$\cdot$mol$^{-1}$ with a Pearson's coefficient of 0.84, the largest difference in absolute value found is 4.3 kcal$\cdot$mol$^{-1}$. For the molecules studied here the MAE of the difference of errors of the methods is 7.7 kcal$\cdot$mol$^{-1}$, while the largest difference in absolute value is 15.9 kcal$\cdot$mol$^{-1}$.

As said before, the isodesmic scheme was employed to provide better theoretical estimates. Table \ref{table isodesmicas} contains a summary of the results. At a glance, the difference among the values for each molecule have declined dramatically as it was expected, which implies that large error compensations have taken place. The only molecule for which experimental values exist is ClF$_5$, it is around -57.2 kcal$\cdot$mol$^{-1}$ as discussed in Section \ref{sec ClF3 y ClF7}. The current best calculation for this molecule was performed at the CCSD(T)/CBS level of theory using atomization scheme, giving -54.5  kcal$\cdot$mol$^{-1}$\cite{doi:10.1021/ic301438b}. It is noteworthy that every value in Table \ref{table isodesmicas} for ClF$_5$ exceed the accuracy of that result.

\subsection{Multiconfigurational character \label{sec multiconfigurational diagnostic}}

For practical purposes the correlation energy is artificially and loosely divided in two contributions, dynamical correlation and static or nondynamical correlation. This distinction is rather arbitrary, physically speaking both arise from Coulomb interactions and from a mathematiacal perspective both can be obtained with linear combinations of Slater determinants\cite{tew2007electron}. However, this distinction can be useful to guide a theoretical treatments. 

There are many diagnostic methods to quantify the importance of nondynamical correlation\cite{fogueri2013simple}. In this work the widely used $T_1$ diagnostic method\cite{lee1989diagnostic} was tried, together with the \%TAEe[T]\cite{karton2011w4-11} method and the new $A_{25\%}$[PBE] and $A_{25\%}$[BLYP]\cite{fogueri2013simple} tests. The DFT calculations were carried out with Orca 2.9.1 software\cite{neese2012orca}, with the def2-QZVPP\cite{weigend2005balanced} basis set which assure near CBS values, using B3LYP/def2\-TZVPP\cite{b3lypcita,weigend2005balanced} equilibrium geometries. Test results are presented in Table \ref{table diagnosticos multireferencia}

\begin{table}[htbp]
	\begin{center}
		\begin{tabular}{lccc}
			\hline
			&$T_1$	&$A_{25\%}$[BLYP]	&$A_{25\%}$[PBE]\\
			\hline
			ClF	&0.011	&0.62	&0.62\\
			ClF$_2$	&0.043	&1.04	&1.02\\
			ClF$_3$	&0.017	&0.89	&0.88\\
			ClF$_4$	&0.029	&1.10	&1.07\\
			ClF$_5$	&0.017	&1.00	&0.96\\
			ClF$_6$	&0.020	&1.15	&1.10\\
			ClF$_7$	&0.015	&1.56	&1.41\\
			Cl$_2$F	&0.040	&1.01	&0.98\\
			Cl$_3$F$_2$	&0.034	&1.17	&1.11\\
				\hline
		\end{tabular} 
		\caption{Results of nondynamical correlation diagnostics.\label{table diagnosticos multireferencia}}
	\end{center}
\end{table}

It was suggested that $T_1>0.02$ probably indicates multireference behavior\cite{lee1989diagnostic}. It is not necessary true for open shell systems\cite{henry2002}. This test turns to be useless for predicting the multireference character of these species, as values below 0.02 for closed shell species do not imply single reference character\cite{doi:10.1021/ct2006852}. On the other hand, $A_{25\%}$[BLYP]	and $A_{25\%}$[PBE] tests provide useful results. According to the authors\cite{fogueri2013simple} ``{A$_k$ values around or above 1 appear to indicate severe-to-pathological static correlation. Values around 0.5 appear to indicate moderate-to-severe nondynamical correlation. Values near 0.3 appear to indicate moderate nondynamical correlation. Near 0.15: mild. Below about 0.10: correlation is primarily dynamic in character''.  According to this and Table \ref{table diagnosticos multireferencia}, the A$_\lambda$ diagnostic suggests that the molecules would present severe to pathological static correlation.

For the triatomic species ClF$_2$ and Cl$_2$F, \%TAEe[T]\cite{karton2011w4-11} diagnostic values were calculated with the G09 program using the aug-cc-pV(5+d)Z basis set, giving 14.5\% and 17.3\% respectively. It was necessary to use smaller basis sets for larger compounds. With exception of ClF$_4$, they were calculated by extrapolation using the values from def2-SVP\cite{weigend2005balanced} and def2-TZVP basis set in conjunction with the Helgaker's extrapolation method. The PSI4 program\cite{turney2012psi4} was used for that purpose. Due to convergence difficulties, the reported ClF$_4$ value corresponds to the def2-SVP basis set. It was obtained 13.5, 21.8, 15.9, 21.0, 29.9 and 22.2\% for ClF$_3$, ClF$_4$, ClF$_5$, ClF$_6$, ClF$7$ and Cl$_3$F$_2$ respectively. According to the authors of the test ``below 2\% indicates systems dominated by dynamical correlation; 2-5\% mild nondynamical correlation; 5-10\% moderate nondynamical correlation; and in excess of 10\% severe nondynamical correlation.''\cite{karton2011w4-11}.  Therefore, the  \%TAEe[T] suggests severe nondynamical correlation for the studied compounds.

Unfortunately, anomalous large basis set dependence is found in the value of \%TAEe[T] for ClF$_7$. The raw def2-SVP and def2-TZVP values obtained are 69.3\% and \%34.0. This remarkable, because authors of \%TAEe[T] diagnostic found very little basis set dependence in the 140 molecules that they tested. Their double zeta values are generally within 1\% of the basis set limit values, and they found that for pathologically multireference cases the deviation may reach up to 2\%. Although they used little larger basis sets, these differences must be due to the ClF$_7$ behavior. The value obtained through extrapolation (29.9\%) is assumed to be close to the basis set limit. Taking advantage of the fast convergence of (T) contribution to atomization energy\cite{ranasinghe:144104}, by using cc-pV(D,T)Z basis set extrapolation with Helgaker's method\cite{helgaker1997basis} and the energy values reported by Dixon\cite{doi:10.1021/ic301438b}, a rude estimation of the \%TAEe[T] was done, resulting in $\approx$30\%. This coincidence supports the reliability of the previous value.

Results of the test should not be surprising nor conclusive. The \%TAEe[T] test is founded on the approximate linear relationship between \%TAEe[T] and the percentage in which superior excitations contribute to the total atomization energies. The results of the \%TAEe[T] diagnostic suggest that, for the molecules studied,  an uncommon large percentage of the TAE may correspond to the higher excitations. As A$_\lambda$ is also approximately proportional to \%TAEe[T]\cite{fogueri2013simple}, its results allows to draw similar conclusions. Both diagnostic are only indirectly connected with static correlation, however their results are very useful here in that they support the suspicions about similar tendencies on the behavior of species without experimental values.

It is interesting to explore the particular reasons for this behavior because, as said before, it is expected that the importance of higher excitations increase in multireference systems. A more direct approach is to inspect the orbital occupancies of a multi-configurational self-consistent field  calculation. It was done for the four closed shell species. A stability analysis performed with Gaussian 09 did not reveal instabilities. Complete Active Space SCF (CASSCF) calculations\cite{schlegel1982mc,bernardi1984mcscf,frisch1992evaluation,yamamoto1996direct,hegarty1979application,eade1981direct} were performed employing ORCA using natural orbitals\cite{lowdin1956natural} coming from previous MP2/def2-TZVP\cite{moller1934note,head1988mp2}  calculations. To correlate the entire valence space is prohibitively expensive except for ClF. Smaller spaces were explored, in Figure \ref{fig:NOON} orbital occupancies were plotted for the active spaces considered. Calculations using other active spaces were performed, they are not qualitatively different for the purpose, so details are omitted to conserve space. 

\begin{figure}[htbp]
	\centering
	\includegraphics[width=\columnwidth]{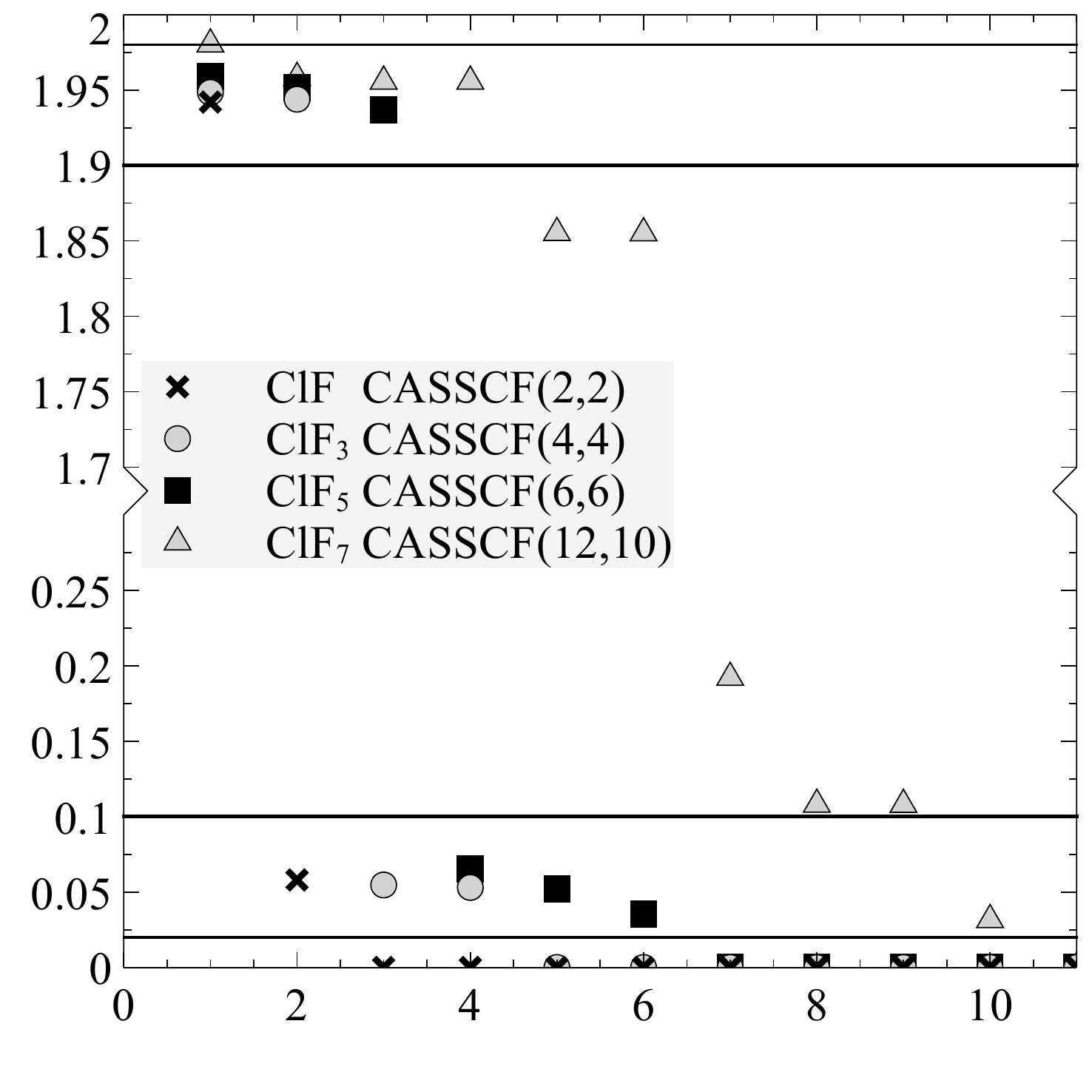}
	\caption{Occupancy of orbital $n$ vs. $n$, starting from $n=1$ for the first (lowest energy) orbital from the active space.}
	\label{fig:NOON}
\end{figure}

The active spaces were sought trying to avoid the inclusion of orbitals with close to empty ($\lesssim$ 0.02) or  full ($\gtrsim 1.98$) occupancies. The orbitals of ClF$_7$ with occupancies between 0.1 and 1.9 indicate that this specie has considerable multiconfigurational character\cite{schmidt1998construction}. The remaining active orbitals can be considered borderline cases\cite{keller2015selection}. This suggests that the use of methodologies with good error compensation becomes more important. The methods that add dynamical correlation on top of a multicomfigurational calculation are prohibitively expensive for these cases, in particular if one takes into consideration that the dynamical contribution does not seem to be small\cite{pettersson1983binding}.  Because of this, it is convenient to estimate the values using a good single reference method in a error compensating scheme like the ones used in this work.

\subsection{The ClF$_3$ and ClF$_7$ cases \label{sec ClF3 y ClF7}}

A special situation arises in the case of ClF$_7$ molecule, where the largest differences between the values of the methods arise to 25.8 kcal$\cdot$mol$^{-1}$ and 7.2 kcal$\cdot$mol$^{-1}$, for the atomization and isodesmic reaction schemes respectively. BAC procedures, especially BAC-G3B3, showed high accuracy for the three closed shell experimentally characterized species. At first glance, BAC procedures should also be reliable for ClF$_7$  if there are similar chemical environments for F-Cl bonds in all these molecules. Although Mulliken atomic charges\cite{mulliken:1833} and Mayer Bond orders\cite{Mayer1983270} have reasonable similarities in ClF$_3$, ClF$_5$ and ClF$_7$, according to orbital occupancies and diagnostic it is  not the case. The approximated mean energies per mol of F-Cl bonds of the closed shell series are -13.3 , -13.1, -11.4 and 2 kcal$\cdot$mol$^{-1}$ so the good performance BAC methods in principle cannot be extrapolated to ClF$_7$.

The error in the heat of formation predicted by using an isodesmic reaction is due to the errors in the calculated heat of reaction and due to the experimental errors. The former can be greatly reduced if good error compensation takes place. The similarity between the ClF$_n$ ($n$=1,3,5) suggest that there is a very good  error compensation in the isodesmic reaction used for ClF$_5$. If such is the case, it is convenient to use this reaction for estimate the formation enthalpy of ClF$_3$ because the coefficients for experimental values of ClF and ClF$_5$ are divided by two, decreasing the effect of experimental uncertainty. By performing the corresponding calculation is seen that all the composite methods differ from each other in less than 0.2 kcal$\cdot$mol$^{-1}$ for the reaction energy. These results motivated the calculation of this reaction energy using the values calculated at CCSD(T)/aug-cc-pV(5+d)Z level of theory reported in ref. \cite{doi:10.1021/ic301438b}, which is more robust and it is expected to bring better error compensation. It gives a value that only differs in 0.1 kcal$\cdot$mol$^{-1}$ from the composite methods used in this work. This coincidence among methods strongly suggests that the reaction energy was obtained very accurately.

The experimental ClF$_5$ value  with lower uncertainty is -56.9$\pm$~1.7~kcal$\cdot$mol$^{-1}$ \cite{gurvichtables1994}. This implies that in absence of errors due to the calculation of the reaction energy, the calculated heat of formation of ClF$_3$ has less uncertainty than the experimental one (-39.3 $\pm$ 1.2 kcal$\cdot$mol$^{-1}$ \cite{gurvichtables1994}).

Other experimental values for the ClF$_5$ molecule are available\cite{doi:10.1021/ic301438b}: -59.9$\pm$3.9~kcal$\cdot$mol$^{-1}$, -57.2$\pm$4.2 kcal$\cdot$mol$^{-1}$, -56$\pm$4.5 kcal$\cdot$mol$^{-1}$. To choose only one value to be employed in the isodesmic scheme calculation without discarding information, the  minimum-variance unbiased estimator of the mean ($\tilde{\theta}$) was taken. It is obtained, under the assumption that the observations are unbiased, by taking a linear combination of the means $\{\theta_i\}$ weighted by the inverse of the associated variances. The estimated variance ($\sigma^{2}(\tilde{\theta})$) is obtained as the inverse of the sum of the inverse of the variances $\{\sigma_i ^2\}$. As the true variances are unknown they were equated to those estimated in the experiments. That is, for the experimental available values $\{(\theta_i,\sigma_i)\}$, the estimations are computed as\cite{hartung2011statistical}

\begin{align}\nonumber
\tilde{\theta} = \frac{\sum_i \sigma^{-2}_i \theta_i}{\sum_i \sigma^{-2}_i } & & \sigma^{2}(\tilde{\theta}) = \frac{1}{\sum_i \sigma^{-2}_i }
\end{align}

 This procedure gives -57.2 $\pm$ 1.4 kcal$\cdot$mol$^{-1}$ for the reference enthalpy of formation of ClF$_5$. Then, neglecting the error for the reaction enthalpy, the estimated formation enthalpy at 298.15 K of ClF$_3$ is -39.75$\pm$ 0.7  kcal$\cdot$mol$^{-1}$. As this result is completely independent of the experimental value of this molecule, they may be combined through the procedure above to get the final value of -39.6$\pm$0.6 ~kcal$\cdot$mol$^{-1}$.

The use CCSD(T)/aug-cc-pV(5+d)Z in the isodesmic reaction scheme together with the accepted values above gives 16.6kcal$\cdot$mol$^{-1}$ for the standard reaction enthalpy of ClF$_7$ at 298.15 K. Another isodesmic and isogyric reaction was also tested to check consistency: 

\begin{table}[htbp]
    \begin{tabular}{lcl}
        ClF$_7$ + ClF$_3$ $\leftrightarrow \,$2 ClF$_5$
    \end{tabular}
\end{table}

The corresponding value is 17.2 kcal$\cdot$mol$^{-1}$ .

\subsection{Recommended Values \label{sec recommended}}

Some compounds (ClF$_2$, ClF$_3$, ClF$_5$ and ClF$_7$) have received a special treatment as discussed above. For the others the best we can do is contemplate the strengths and weaknesses of the methods qualitatively and make a conscious guess. As said before, best estimates must come mostly from isodesmic calculations (or experiments when available), specially from those using G4 model chemistry. Notice, that as the values of the methods are not truly independent of each other, the  minimum-variance unbiased estimator of the mean can not be used for this purpose. Best estimates of this work are summarized in Table \ref{resultados finales}, corresponding values for homolitic elimination of a fluorine atom were also tabulated when they are available from suggested heat of formation.

\begin{table}[htbp]
	\begin{center}
		\begin{tabular}{lcc}
			\hline
			Molecule &$\Delta_f H(298.15K)$ & $\Delta_{dis} H(298.15K)$\\
			\hline
			ClF	 &-13.3 & 61.2 \\
			ClF$_2$	 &-10.7 & 16.4\\
			ClF$_3$	 &-39.6 & 47.9 \\
			ClF$_4$	 &-31.0 & 10.4\\
			ClF$_5$	 &-57.2 & 45.2 \\
			ClF$_6$	 &-41.7 & 3.5\\
			ClF$_7$	 &16.8 & -39.5\\
			Cl$_2$F	 & 11.9& 7.1/3.8\\
			Cl$_3$F$_2$& 18.3 & - \\
			\hline
		\end{tabular} 
		\caption{Estimates for values of standard formation enthalpy, and for the enthalpy change for the homolitic elimination of a fluorine/chlorine atom, at 298.15K in units of kcal$\cdot$mol$^{-1}$.\label{resultados finales}}
	\end{center}
\end{table}

Reliability is very case-dependent. The ClF value is very well established both empirically\cite{atct} and theoretically\cite{karton2011w4-11}. It should be expected that the ClF$_2$ predicted value is to be very accurate, probably with an error significantly lower than about 1~kcal$\cdot$mol$^{-1}$. As these composite methods have not been exhaustively tested on radicals, the values of ClF$_4$, ClF$_6$ and Cl$_2 $F should be considered less reliable and errors of about 2 kcal$\cdot$mol$^{-1}$  should not be surprising. For the ClF$_5$ molecule the estimated uncertainty is about 1~kcal$\cdot$mol$^{-1}$, as the underling assumptions of the statistical treatment can not be proved, but it is in good accordance with the experimental value of chlorine trifluoride. The value for ClF$_3$ gives us more confidence and makes us expect an error lower than about 1~kcal$\cdot$mol$^{-1}$. For the difficult ClF$_7$ case we must be more conservative and estimate an error range not smaller than 2~kcal$\cdot$mol$^{-1}$.

\subsection{Comparison with previous results \label{sec comparison previous results}}

As far as is known in this work there are only two other works (refs. \cite{ricca2000heats,doi:10.1021/ic301438b}) that employed equal or higher level of theory on some of the compounds in question. In one of them\cite{doi:10.1021/ic301438b}, authors concluded that the closed shell compounds here are monodeterminantal cases because the results of the $T_1$ test, and they estimate error bars of $\pm$ 1.5 kcal$\cdot$mol$^{-1}$ for their computed heats of formation, calculated via total atomization scheme. In that work, formation enthalpies were calculated  with two methods which are very similar to each other. For the one that uses larger basis sets, they got the values -13.1 kcal$\cdot$mol$^{-1}$, -38.3 kcal$\cdot$mol$^{-1}$, -54.5  kcal$\cdot$mol$^{-1}$ and  20.7 kcal$\cdot$mol$^{-1}$, for ClF, ClF$_3$, ClF$_5$ and ClF$_7$ respectively.

Recommended values in the present work have some discrepancies with those mentioned above, especially for the ClF$_7$ case, for which a value around 16.8 kcal$\cdot$mol$^{-1}$ is suggested (see Table  \ref{resultados finales}).  That is, values differ in almost 4 kcal$\cdot$mol$^{-1}$. 

For ClF$_3$, the raw value from this work is -39.75$\pm$ 0.7 kcal$\cdot$mol$^{-1}$. Again, this is lower than the corresponding result form reference \cite{doi:10.1021/ic301438b}: -38.3 kcal$\cdot$mol$^{-1}$. It is convenient to sound a note of caution: The experimental value -38.0 $\pm$ 0.7 kcal$\cdot$mol$^{-1}$ from NIST-JANAF \cite{1998nist}  should not be used for comparisons. This value was criticized by Ricca in favour of the Gurvich's one (-39.3 $\pm$ 1.2 kcal$\cdot$mol$^{-1}$ \cite{gurvichtables1994}). The latter was then adopted by NIST’s "Computational Chemistry Comparison and Benchmark DataBase"\cite{johnson1998nist}. It is interesting to note that the value -38.0 kcal$\cdot$mol$^{-1}$ was obtained indirectly by using the enthalpy of the thermal decomposition reaction of ClF$_3$, giving ClF and F$_2$ as products, and using a value of -12.0 kcal$\cdot$mol$^{-1}$ for the standard enthalpy of formation of ClF. Again, this value was criticized by Ricca in favour of the newer Gurvich's one, that is, -13.3 $\pm$ 0.1 kcal$\cdot$mol$^{-1}$.  The latter coincides with the Active Thermochemical Tables\cite{atct} -13.308 $\pm$ 0.014 kcal$\cdot$mol$^{-1}$. The highest level calculation performed on the ClF molecule is that by Karton\cite{karton2011w4-11}, which gives an atomization value that, once converted to standard formation enthalpy at 0 K, is at 0.1 kcal$\cdot$mol$^{-1}$ from the experimental value reported in the reference \cite{atct}. If the newer and more reliable value of ClF molecule is used, the old NIST-JANAF value for ClF$_3$ will result in coincidence with the Gurvich’s value.

If values of this work are compared to those from reference \cite{doi:10.1021/ic301438b}, there is a clear tendency to get lower values as molecular mass increases. Differences may lie in two factors: in this series of compounds basis set requirements may increase with the number of fluorine atoms, as has been claimed before \cite{ricca2000heats} for ClF, ClF$_2$ and ClF$_3$; and the error of their result seems to increase with the number of Cl--F bonds.

Large differences as in the ClF$_7$ are reasonable for multi-reference cases. We can see it by using  Figure \ref{fig:tae} where the contribution (in percentages) of excitations higher than (T) to the atomization energy is plotted vs. the corresponding contribution of perturbative triples excitations, for the results published in \cite{karton2011w4-11}. The atomization energy for this molecule rounds about 150-160 kcal$\cdot$mol$^{-1}$, and its \%TAEe[T] $\approx$30\%, so employing Figure \ref{fig:tae} an error of about 4 kcal$\cdot$mol$^{-1}$ for the CCSD(T) method is fairly plausible. 

It is very interesting to note that it is possible to have a good idea of the magnitude of the error bounds of the contribution of higher excitations to the formation enthalpy when the \%TAEe[T] values are not too large. That is to say, for CCSD(T)/CBS calculations when mild static correlation is present (\%TAEe[T] $<$ 4\%), a good estimate of the error bounds in heats of formation calculated via atomization reactions is given by $\approx 0.3\%$ of the total atomization energy.

\begin{figure}[htbp]
\centering
\includegraphics[width=\columnwidth]{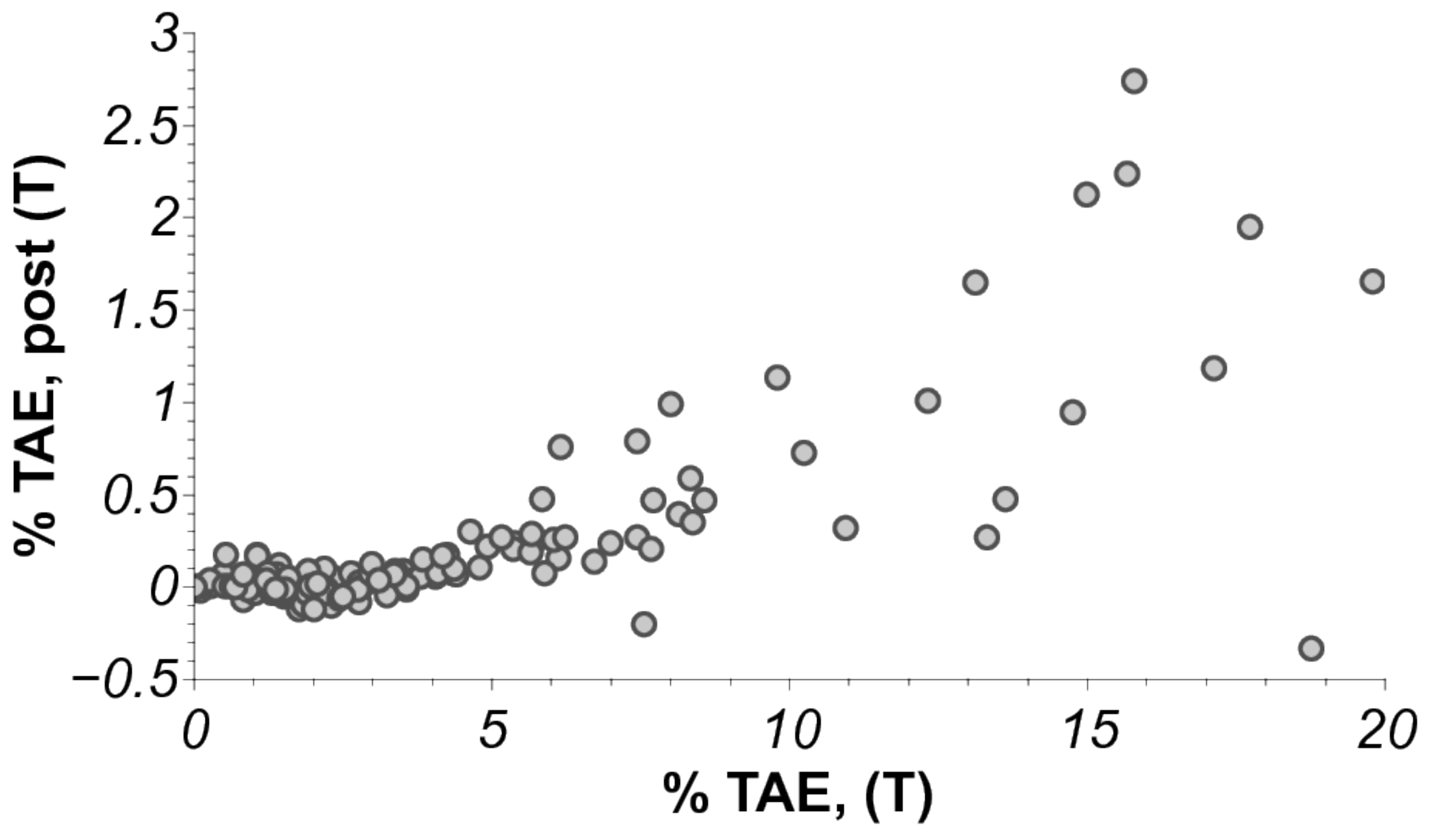}
\caption{Percentage in which above (T) excitations contributes to the total atomization energy vs. the corresponding percentage for (T). }
\label{fig:tae}
\end{figure}

With the help of the results published in ref. \cite{karton2011w4-11}, it is simple to get the errors of the estimated TAE at 0K for F$_2$ and Cl$_2$ at the CCSD(T)/CBS level of theory.  By using  the corresponding multiplicative factors (due to stoichiometry) they can be added to the formation enthalpies calculated in the ref.  \cite{doi:10.1021/ic301438b}, to get the formation enthalpies calculated through the formation reactions. This procedure implies to discard the negligible differences due to errors in thermal corrections and gives -13.6 kcal$\cdot$mol$^{-1}$, -39.6  kcal$\cdot$mol$^{-1}$, -56.5  kcal$\cdot$mol$^{-1}$ and 17.9  kcal$\cdot$mol$^{-1}$ for ClF, ClF$_3$, ClF$_5$ and ClF$_7$ respectively. It can be seen that the values for ClF$_3$, ClF$_5$ and ClF$_7$ are now closer to those recommended in the present work. In fact for the ClF$_3$ molecule the values are exactly the same.  The tendency to obtain lower values, in the present work, as the fluorine atoms number increases can still be seen, and it can be argued that this is due to the better error compensations  that takes place in isodesmic reactions. The ClF value predicted by this procedure is lower than the experimental one in 0.3 kcal$\cdot$mol$^{-1}$, this is expected due to the relatively large contribution to energy of  ''post (T) excitations" in the difluorine molecule.  According to the values reported in ref.  \cite{karton2011w4-11} the post (T) contribution to the formation reaction is 0.25 kcal$\cdot$mol$^{-1}$. The comments in the current paragraph  support the main hypothesis and the recommended values from the present work.

\subsection{Standard thermodynamic functions \label{sec standard thermodynamic functions}}
Standard heat capacities at constant pressure and standard entropies were calculated based on statistical thermodynamics at temperatures between 200 and 800 K. The B3LYP/6-311+G(3d2f) level of theory was used for the calculation of rotational and vibrational constants. All usual approximation were used, i.e. separability of movements contributions, ideal gas law behavior, replacement of summation over the discrete levels by integration over continuous levels when feasible, neglect of degeneracies to nuclear spins   etc.. Thermal changes in enthalpy were also computed and reported as the difference between its value at the temperature $T$ and the one corresponding at 298.15 K. The results are presented in Table \ref{table termodinamica}.

\begin{table*}[]
	\begin{center}
		\begin{tabular}{crrrrrrrr}\hline
		Specie &  Property &200 K &298.15 K &400 K& 500 K& 600 K& 700 K& 800 K \\ \hline
	&$H^\circ_m(T^*)$	&-3.0 &0.0	&3.3	&6.8	&10.3	&13.8	&17.4\\
ClF	&$C^\circ_{p,m}(T)$	&30.1	&32.0	&33.6	&34.6	&35.4	&35.8	&36.2\\
	&$S^\circ_m(T)$	&205.4	&217.7	&227.4	&235.0	&241.4	&246.8	&251.7\\\hline
	&$H^\circ_m(T^*)$	&-4.7	&0.0	&5.3	&10.7	&16.2	&21.8	&27.5\\
ClF$_2$	&$C^\circ_{p,m}(T)$	&45.1	&50.3	&53.2	&54.8	&55.8	&56.4	&56.8\\
	&$S^\circ_m(T)$	&247.1	&266.2	&281.4	&293.4	&303.5	&312.2	&319.7\\\hline
	&$H^\circ_m(T^*)$	&-5.9	&0.0	&7.0	&14.3	&21.9	&29.7	&37.6\\
ClF$_3$	&$C^\circ_{p,m}(T)$	&54.0	&64.8	&71.3	&74.9	&77.2	&78.7	&79.6\\
	&$S^\circ_m(T)$	&265.4	&289.2	&309.2	&325.6	&339.5	&351.5	&362.0\\\hline
	&$H^\circ_m(T^*)$	&-7.8	&0.0	&9.2	&18.9	&28.9	&39.1	&49.4\\
ClF$_4$	&$C^\circ_{p,m}(T)$	&70.9	&85.9	&94.1	&98.6	&101.3	&103.0	&104.1\\
	&$S^\circ_m(T)$	&286.7	&318.1	&344.6	&366.1	&384.3	&400.1	&413.9\\\hline
	&$H^\circ_m(T^*)$	&-8.7	&0.0	&10.7	&22.2	&34.2	&46.5	&59.1\\
ClF$_5$	&$C^\circ_{p,m}(T)$	&76.9	&98.4	&110.8	&117.7	&122.0	&124.7	&126.5\\
	&$S^\circ_m(T)$	&289.5	&324.6	&355.5	&381.0	&402.9	&421.9	&438.7\\\hline
	&$H^\circ_m(T^*)$	&-11.4	&0.0	&13.5	&27.6	&42.2	&57.1	&72.2\\
ClF$_6$	&$C^\circ_{p,m}(T)$	&103.7	&125.5	&137.4	&144.0	&147.9	&150.4	&152.1\\
	&$S^\circ_m(T)$	&319.4	&365.4	&404.1	&435.5	&462.1	&485.1	&505.3\\\hline
	&$H^\circ_m(T^*)$	&-11.6	&0.0	&14.4	&30.0	&46.4	&63.2	&80.4\\
ClF$_7$	&$C^\circ_{p,m}(T)$	&100.9	&131.9	&150.1	&160.3	&166.5	&170.6	&173.3\\
	&$S^\circ_m(T)$	&317.9	&364.5	&406.1	&440.8	&470.6	&496.6	&519.5\\\hline
	&$H^\circ_m(T^*)$	&-5.1	&0.0	&5.6	&11.2	&16.8	&22.5	&28.3\\
Cl$_2$F	&$C^\circ_{p,m}(T)$	&49.5	&53.5	&55.4	&56.3	&56.9	&57.2	&57.4\\
	&$S^\circ_m(T)$	&261.1	&281.8	&297.8	&310.3	&320.6	&329.4	&337.0\\\hline
	&$H^\circ_m(T^*)$	&-8.8	&0.0	&9.9	&20.0	&30.4	&40.8	&51.4\\
Cl$_3$F$_2$	&$C^\circ_{p,m}(T)$	&84.1	&94.5	&99.7	&102.4	&104.1	&105.1	&105.8\\
	&$S^\circ_m(T)$	&323.5	&359.3	&387.8	&410.4	&429.2	&445.4	&459.4\\\hline
		\end{tabular} 
		\caption{Standard molar entropies ($S^\circ_{m}(T)$ in $JK^{-1}$mol$^{-1}$) and standard molar heat capacities at constant pressure ($C^\circ_{p,m}(T)$ in $JK^{-1}$mol$^{-1}$), difference between molar standard enthalpy at temperature $T$ and  molar standard enthalpy at 298.15 K ($H^\circ_m(T^*)$ in $kJ\cdot$mol$^{-1}$).\label{table termodinamica}}
	\end{center}
\end{table*}
\normalsize

Calculated $S^\circ_{m}(T)$ and $C^\circ_{p,m}(T)$ values are in reasonable accordance with the experimental ones and with those reported in ref. \cite{du2011theoretical}. The last ones, obtained at HF/6-31G(d) level of theory, were recalculated in this work and some noticeable differences were found for many cases. The new values, at 298.15 K, agree exactly with those reported in the thermo-chemistry analysis provided by the Gaussian 09 software. In order to analyze the importance of the basis set effect, the calculations were also performed using the larger basis set 6-311+G(3d2f). Results do not differ significantly from those obtained with 6-31G(d). The detailed values are available in the Supporting Information.

For completeness, to be able to work at any temperature in the range 200-800 K, three well known functions were fitted to interpolate the tabulated results. By abuse of language they all are often referred as ``NASA Polynomials'' and they are defined by the equations

\begin{align}\nonumber
H^\circ_{m}(T)/(RT) &:=  \sum _{i=1} ^5  i^{-1} \,a_i\, T^{i-1} +  a_6 T^{-1} \\\nonumber
C^\circ _{p,m}(T)/R &:= \sum _{i=1} ^5  b_i \, T^{i-1} \\\nonumber
S^\circ _m(T) /R  &:= c_1 \ln(T) +  \sum _{i=2} ^5  c_i \, T^{i-1} + c_6\\\nonumber
\end{align}

The obtained coefficients can be found in the Supporting Information. Note that they are reported mostly due to common practice, and that simpler interpolation schemes should suffice.

\subsection{Potential energy curves \label{sec potential energy curves}} 

Correct description of potential energy surfaces (PES) is essential for many chemistry application, especially for kinetics studies where accurate description of energy  curves are usually needed. Some uncommon behavior  were found in the description of the energy elimination curves from chlorine fluorides, throughout the use of mono reference wave function-based methods.  It can be illustrated with the peculiar situation that arises around the equilibrium geometry of the Cl$_2$F radical, in which triples excitations are needed into the couple clusters framework to do not predict a dissociative geometry. It is shown in Figure \ref{fig:ELIMINACIONCL} where elimination curves at B3LYP/def2-TZVPP geometries are plotted, and negative slope indicate dissociative behavior. 

\begin{figure}[htbp]
	\centering
	\includegraphics[width=\columnwidth]{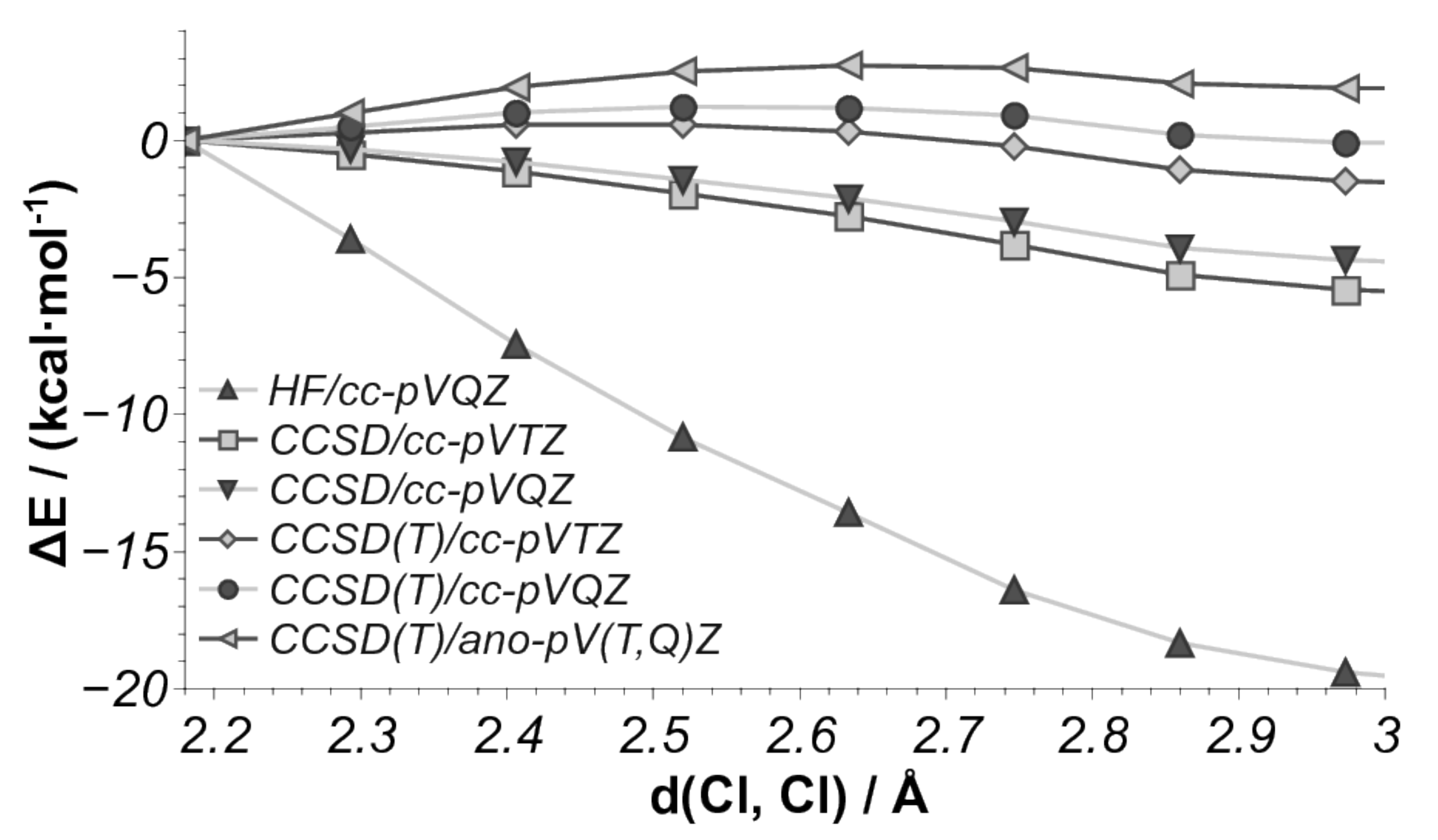}
	\caption{Energy elimination curves of a chlorine atom from Cl$_2$F for HF/cc-pVQZ, CCSD/cc-pVTZ, CCSD/cc-pVQZ, CCSD(T)/cc-pVTZ, CCSD(T)/cc-pVQZ and CCSD(T)/ano-pV(T,Q)Z methods. Geometries where optimized at B3LYP/def2-TZVPP level of theory.}
	\label{fig:ELIMINACIONCL}
\end{figure}

\subsection{Density Functional Theory Calculations \label{sec dft calculations}}
Due to the difficulties in retrieving correlation energy in chlorine fluorides, it is not expected that low correlated ab-initio single reference methods retrieve reliable energy values. BAC procedures performed reasonably well, especially for closed shell cases, but they are parametrized for equilibrium geometries so results are uncertain in other cases. Far from equilibrium geometries, the importance of correlation energy grows. As noted before, very extended basis sets are required when dealing with these species using ab-initio single reference methods. This is due almost exclusively to correlation part because, as it is well known, the convergence of the correlation part is slower than that of the Hartree-Fock method\cite{halkier1999basis}. In brief, for predicting reasonable elimination curves using single reference ab-initio methods, very large basis sets and high correlated methods are necessary, which is very computationally demanding. 

DFT calculations do not suffer this basis sets drawback, and as they present some implicit coverage of multireference effects\cite{cremer2002implicit,polo2002some}, it is interesting to test many functionals for these species. Nevertheless, it has been well tested that DFT results for multi-reference systems are normally poorer than those corresponding to systems with small static correlation\cite{karton2011w4-11,goerigk2010efficient}.

Atomization energies for the chlorine fluorides was computed in this work, using B3LYP/def2-TZVPP geometries, and more than fifty different functionals with the def2-QZVPP basis set (see the Supporting Information for details). Formation enthalpy was calculated adding the difference between formation enthalpy and atomization electronic energy for the G4 method, to the atomization electronic energy obtained for the functionals.

Results of some of the best performers are listed in Table \ref{table best dft}, see the Supporting Information for the complete list and references. The better performance of the M05 functional with the smaller 6-311+G(3d2f) basis set, can be understood in terms of the similarities between this basis set and the one used for the parametrization of the functional.

\begin{table}[hb]
	\begin{center}
		\begin{tabular}{lcccc}
			\hline
			&MAE	&RMSE	&MSE	&LAE	\\
			\hline
			B3LYP	&4.4	&5.7	&0.9	&12.0\\
			PBE0	&3.9	&4.8	&-2.9	&-9.5\\
			PW6B95	&3.6	&4.3	&-2.2	&8.2\\
			RI-B2PLYP	&2.6	&3.1	&0.6	&-5.3\\
			M06	& 2.2	&2.6	&-0.2	&4.1\\			
			RI-PWPB95	&1.9	&2.4	&0.1	&-4.2\\
			M05$^a$	&2.7	&3.5	&0.4	&7.6\\
				\hline
		\end{tabular} 
		\caption{Mean absolute error (MAE), root mean square error (RMSE), mean signed error (MSE) and largest absolute error (LAE), accepted values are those from Table \ref{resultados finales}.  Values are in kcal$\cdot$mol$^{-1}$.\label{table best dft} $^a$Geometry optimization and single point energy calculations using the 6-311+G(3d2f) Pople’s basis set
		}
	\end{center}
\end{table}

Potential energy curves must be carefully analyzed for the functional in question.  In this work, the energy curves for the elimination of an atom of fluorine/chlorine from chloride fluorides, were calculated. They include singlet and triplet states for the dissociation of closed shell species. 
 
 The functional chosen,  belonging to the Minnesota 06 family, was the Truhlar's hybrid M06 functional, because it was the most accurate among of the hybrid functional used here. Double hybrid functionals were deliberately avoided because of its additional complexity, although it can be interesting to test the PWPB95 functional in future works. The omission of ClF$_7$ molecule is because the homolitic elimination of a fluorine atoms is very unfavorable respect of the elimination of a F$_2$ molecule, so the later is the natural elimination. 

Geometry optimization and the single point calculation were performed with the def2-TZVP basis sets, except for the radicals Cl$_2$F and Cl$_3$F$_2$ for which the def2-SVP basis set was used during the geometry optimization step. The Orca software was chosen to perform the calculations, because its ability to use the  chain-of-spheres algorithm\cite{neese2009efficient} that allows to save considerable computation time. The curves can be found in the Supporting Information. It is seen that the M06 method behave qualitatively correct, so its usage is recommended as a  low cost alternative for the matter.

\section{Concluding Remarks}
Good estimated values for formation enthalpies calculated via atomization scheme require to go beyond CCSD(T)/CBS for the heaviest members. Using isodesmic reaction scheme provides very high systematic error compensation which can even include non-calculated higher excitations. Based on this approach, new estimates for standard enthalpies of formation were suggested.

Bond additivity corrections work properly for most chlorine fluorides but they are only reliable for equilibrium geometries where they were parametrized. Composite methods which include low correlated ab-initio calculations should be avoided or used with extreme caution for these cases.

In this work it was found that the \%TAEe[T] diagnostic may show large basis set dependence, because of this it is encouraged the use of A$_k$ diagnostic method for casual users. 

A general method was suggested for the prediction of error bounds on heats of formation calculated via atomization reactions with CCSD(T)/CBS method, for cases with as much as mild multireference character.

For the study of cases where energies corresponding to non-equilibrium geometries are needed, post-Hartree-Fock methods turn to be too expensive. DFT is suggested as a low cost alternative of simple application, through the hybrid M06 functional together with the def2-QZVPP basis sets.

\section*{Acknowledgement}

The author acknowledges the National Scientific and Technical Research Council (CONICET) for the financial support.

\section*{Supporting Information}
Cartesian (x, y, z) coordinates in angstrom of the equilibrium geometries of the compounds. Formation enthalpies via atomization reactions for the considered functionals. NASA polynomials coefficients. Energy elimination curves computed using the M06 functional.

\section*{References}

\bibliography{referencias}

\pagebreak

\widetext

\newgeometry{left=3.00cm, right=3.00cm,top=2cm,bottom=1cm}

\begin{center}
	\textbf{\large Supporting Information: Revisiting the thermochemistry of chlorine fluorides}
\end{center}

\setcounter{equation}{0}
\setcounter{figure}{0}
\setcounter{table}{0}
\setcounter{page}{1}
\makeatletter
\renewcommand{\theequation}{S\arabic{equation}}
\renewcommand{\thefigure}{S\arabic{figure}}
\renewcommand{\bibnumfmt}[1]{[S#1]}
\renewcommand{\citenumfont}[1]{S#1}

\section*{Coefficients calculated of the NASA Polynomials.}

For each molecule the coefficients are given in three list. Those lists contain the unitless ordered $a_i$, $b_i$ and $c_i$ coefficients according to the main text.

\vspace{0.5cm}

ClF
\begin{lstlisting}
[  2.91293550e+00   4.02365992e-03  -2.66532030e-06  -1.71050947e-09
1.96515519e-12   6.10888166e+02]

[  3.04744504e+00   2.95102141e-03   3.24789831e-07  -5.18934744e-09
3.39787945e-12]

[  2.92745774e+00   3.89817221e-03  -2.27603269e-06  -2.22670757e-09
2.21302951e-12   8.46094981e+00]
\end{lstlisting}
\vspace{0.3cm}

ClF$_2$
\begin{lstlisting}
[  2.62362493e+00   2.13117993e-02  -4.48665378e-05   4.49646859e-08
-1.74831921e-11   1.02096815e+03]

[  2.61359898e+00   2.12906699e-02  -4.44166519e-05   4.38304356e-08
-1.66802175e-11]

[  2.61289776e+00   2.14085490e-02  -4.51780418e-05   4.53914796e-08
-1.76941947e-11   1.23825197e+01]
\end{lstlisting}
\vspace{0.3cm}

ClF$_3$
\begin{lstlisting}
[  9.16990494e-01   4.16622427e-02  -8.38849666e-05   8.13047461e-08
-3.08220221e-11   2.29613501e+03]

[  9.30941998e-01   4.13789225e-02  -8.24347833e-05   7.85896173e-08
-2.91397041e-11]

[  9.03014298e-01   4.17895886e-02  -8.42985866e-05   8.18757843e-08
-3.11062406e-11   2.02631676e+01]
\end{lstlisting}
\vspace{0.3cm}

ClF$_4$
\begin{lstlisting}
[  4.98875336e-01   6.16571478e-02  -1.31334000e-04   1.32960984e-07
-5.21387781e-11   2.88169116e+03]

[  3.53349995e-01   6.25385107e-02  -1.32718552e-04   1.32901894e-07
-5.11977574e-11]

[  4.57822363e-01   6.20226988e-02  -1.32498478e-04   1.34542084e-07
-5.29143959e-11   2.19591250e+01]
\end{lstlisting}
\vspace{0.3cm}
\newpage

ClF$_5$
\begin{lstlisting}
[ -1.93174885e+00   8.44276865e-02  -1.73661383e-04   1.71235655e-07
-6.58116987e-11   4.39497872e+03]

[ -1.99074412e+00   8.45374535e-02  -1.72581807e-04   1.67822047e-07
-6.32215025e-11]

[ -1.97078823e+00   8.47788946e-02  -1.74790012e-04   1.72779746e-07
-6.65742198e-11   3.13705501e+01]
\end{lstlisting}
\vspace{0.3cm}

ClF$_6$
\begin{lstlisting}
[  8.17681106e-01   8.96001905e-02  -1.91145824e-04   1.93979901e-07
-7.62624111e-11   4.21711207e+03]

[  4.71977840e-01   9.19792168e-02  -1.96341382e-04   1.97799221e-07
-7.66168483e-11]

[  7.48736929e-01   9.02097757e-02  -1.93076206e-04   1.96587710e-07
-7.75361029e-11   1.97792499e+01]
\end{lstlisting}
\vspace{0.3cm}

ClF$_7$
\begin{lstlisting}
[ -3.95941163e+00   1.21031743e-01  -2.46934732e-04   2.41867391e-07
-9.24548341e-11   6.43653755e+03]

[ -3.94652921e+00   1.20406507e-01  -2.43178348e-04   2.34357029e-07
-8.76368633e-11]

[ -4.00646027e+00   1.21458715e-01  -2.48316743e-04   2.43769611e-07
-9.33990812e-11   3.95227276e+01]
\end{lstlisting}
\vspace{0.3cm}

Cl$_2$F
\begin{lstlisting}
[  3.68626897e+00   1.81936001e-02  -4.18656357e-05   4.48483279e-08
-1.83426669e-11   6.57418637e+02]

[  3.51136536e+00   1.95089423e-02  -4.52323310e-05   4.83019610e-08
-1.95117150e-11]

[  3.66108241e+00   1.84140528e-02  -4.25575664e-05   4.57757205e-08
-1.87924192e-11   9.06273089e+00]
\end{lstlisting}
\vspace{0.3cm}

Cl$_3$F$_2$
\begin{lstlisting}
[  4.37777751e+00   4.51303196e-02  -1.00575758e-04   1.05377832e-07
-4.24219273e-11   1.84079875e+03]
[  4.05937285e+00   4.74746032e-02  -1.06372921e-04   1.10981476e-07
-4.41055544e-11]
[  4.32759990e+00   4.55709056e-02  -1.01962499e-04   1.07241140e-07
-4.33276098e-11   8.63488194e+00]
\end{lstlisting}

\cleardoublepage

\newpage

\section*{Thermodynamics functions obtained by using Hartree-Fock method}

\textbf{Values calculated at  HF/6-31G(d) level of theory.}

\begin{table*}[hb]
	\begin{center}
		\begin{tabular}{crrrrrrrr}\hline
			Specie &  Property &200 K &298.15 K &400 K& 500 K& 600 K& 700 K& 800 K \\ \hline
			&$H^\circ_m(T^*)$	& -3.0 & 0.0 & 3.3 & 6.6 & 10.0 & 13.5 & 17.1 \\
			ClF	&$C^\circ_{p,m}(T)$	& 29.6&31.1&32.7&33.9&34.8&35.3&35.8\\
			&$S^\circ_m(T)$	&205.0&217.0&226.4&233.8&240.1&245.5&250.3\\\hline
			&$H^\circ_m(T^*)$	&-5.3&0.0&6.5&13.4&20.7&28.3&36.1 \\
			ClF$_3$	&$C^\circ_{p,m}(T)$	&48.4&59.5&67.0&71.7&74.7&76.7&78.0\\
			&$S^\circ_m(T)$	&261.0&282.5&301.1&316.6&330.0&341.7&352.0\\\hline
			&$H^\circ_m(T^*)$	&-7.4&0.0&9.6&20.2&31.6&43.4&55.6\\
			ClF$_5$	&$C^\circ_{p,m}(T)$	&64.1&86.1&101.2&110.4&116.3&120.3&123.0\\
			&$S^\circ_m(T)$	&280.1&310.0&337.6&361.2&381.9&400.2&416.4\\\hline
		\end{tabular} 
		\caption{Standard molar entropies ($S^\circ_{m}(T)$ in $JK^{-1}$mol$^{-1}$) and standard molar heat capacities at constant pressure ($C^\circ_{p,m}(T)$ in $JK^{-1}$mol$^{-1}$), difference between molar standard enthalpy at temperature $T$ and  molar standard enthalpy at 298.15 K ($H^\circ_m(T^*)$ in $kJ\cdot$mol$^{-1}$).}
	\end{center}
\end{table*}

\noindent
\textbf{Values calculated at  HF/6-311+G(3d2f) level of theory.}

\begin{table*}[hb]
	\begin{center}
		\begin{tabular}{crrrrrrrr}\hline
			Specie &  Property &200 K &298.15 K &400 K& 500 K& 600 K& 700 K& 800 K \\ \hline
			&$H^\circ_m(T^*)$	& -3.0 &0.0&3.3&6.6&10.0&13.5&17.1\\
			ClF	&$C^\circ_{p,m}(T)$	& 29.6&31.1&32.7&33.9&34.8&35.4&35.8\\
			&$S^\circ_m(T)$	&204.7&216.8&226.1&233.6&239.9&245.3&250.0\\\hline
			&$H^\circ_m(T^*)$	& -5.4&0.0&6.5&13.5&20.9&28.5&36.2\\
			ClF$_3$	&$C^\circ_{p,m}(T)$	&48.6&59.9&67.5&72.1&75.0&76.9&78.3\\
			&$S^\circ_m(T)$	&260.5&282.1&300.9&316.5&329.9&341.6&352.0\\\hline
			&$H^\circ_m(T^*)$	&-7.4&0.0&9.6&20.2&31.6&43.5&55.7\\
			ClF$_5$	&$C^\circ_{p,m}(T)$	&63.2&86.0&101.3&110.7&116.6&120.5&123.2\\
			&$S^\circ_m(T)$	&278.7&308.5&336.1&359.8&380.5&398.8&415.1\\\hline
		\end{tabular} 
		\caption{Standard molar entropies ($S^\circ_{m}(T)$ in $JK^{-1}$mol$^{-1}$) and standard molar heat capacities at constant pressure ($C^\circ_{p,m}(T)$ in $JK^{-1}$mol$^{-1}$), difference between molar standard enthalpy at temperature $T$ and  molar standard enthalpy at 298.15 K ($H^\circ_m(T^*)$ in $kJ\cdot$mol$^{-1}$).}
	\end{center}
\end{table*}

\cleardoublepage

\newpage

\section*{Standard entalphy of formation at 298.15 K in kcal$\cdot$mol$^{-1}$}

Values were calculated by adding to the TAE obtained with the corresponding functional, the
difference between the heat of formation and the TAE for the G4 level of theory.

\begin{table*}[!htb]\footnotesize
	\begin{center}
		\begin{tabular}{crrrrrrrrr}\hline
			&ClF	&ClF$_2$	&ClF$_3$	&ClF$_4$	&ClF$_5$	&ClF$_6$	&ClF$_7$	&Cl$_2$F	&Cl$_3$F$_2$\\\rowcolor{colororca2}\hline
			B1LYP	&-7.2	&-7.6	&-27.5	&-23.9	&-32.5	&-20.6	&52.8	&14.5	&30.5\\\rowcolor{colororca}
			B1P	&-10.1	&-12.5	&-36.6	&-35.6	&-49.8	&-39.8	&28.9	&8.5	&18.7\\\rowcolor{colororca2}
			B2GP-PLYP	&-10.0	&-7.2	&-34.1	&-28.2	&-46.4	&-33.2	&30.4	&16.0	&27.3\\\rowcolor{colororca}
			B2K-PLYP	&-9.3	&-4.1	&-31.5	&-23.2	&-42.2	&-26.6	&36.6	&18.9	&31.6\\\rowcolor{colororca2}
			B2T-PLYP	&-10.0	&-8.3	&-34.2	&-29.1	&-46.0	&-33.0	&31.6	&15.0	&26.8\\\rowcolor{colororca}
			B3LYP	&-10.8	&-15.2	&-38.1	&-38.5	&-50.0	&-41.8	&28.8	&8.2	&18.6\\\rowcolor{colororca2}
			B3P	&-13.1	&-19.2	&-45.5	&-48.1	&-64.1	&-57.6	&9.3	&3.3	&8.9\\\rowcolor{colororca}
			B3PW	&-12.0	&-16.5	&-41.5	&-42.4	&-56.7	&-48.5	&20.1	&5.4	&13.6\\\rowcolor{colororca2}
			BLYP	&-17.6	&-34.4	&-61.1	&-73.7	&-87.1	&-89.5	&-22.3	&-7.5	&-10.5\\\rowcolor{colororca}
			BP86	&-20.6	&-39.6	&-70.6	&-86.0	&-105.2	&-109.7	&-47.4	&-13.6	&-23.0\\\rowcolor{colororca2}
			G1LYP	&-6.4	&	&-24.5	&	&-26.9	&	&61.7	&	&\\\rowcolor{colororca}
			G1P	&-9.3	&-10.3	&-33.4	&	&-43.9	&	&38.3	&	&\\\rowcolor{colororca2}
			G3P	&-12.4	&-17.1	&-42.5	&	&-58.5	&	&18.1	&	&\\\rowcolor{colororca}
			GLYP	&-16.6	&	&-57.2	&	&-80.0	&	&-11.2	&	&\\\rowcolor{colororca2}
			GP	&-19.7	&-36.8	&-66.6	&	&-97.7	&	&-35.6	&	&\\\rowcolor{colororca}
			mPW1LYP	&-8.5	&-10.2	&-31.7	&	&-40.0	&	&42.0	&	&\\\rowcolor{colororca2}
			mPW1PW	&-10.0	&	&-35.9	&	&-48.3	&	&31.4	&	&\\\rowcolor{colororca}
			mPW2PLYP	&-10.1	&-9.5	&-35.0	&	&-47.1	&	&30.1	&	&\\\rowcolor{colororca2}
			mPWLYP	&-19.4	&	&-66.6	&	&-96.9	&	&-36.7	&	&\\\rowcolor{colororca}
			mPWPW	&-21.0	&	&-71.3	&	&-106.1	&	&-48.5	&	&\\\rowcolor{colororca2}
			O3LYP	&-22.6	&-37.4	&-72.7	&-84.2	&-109.7	&-111.1	&-52.1	&-11.4	&-19.3\\\rowcolor{colororca}
			OLYP	&-16.7	&-30.2	&-57.0	&-66.6	&-81.2	&-79.5	&-9.7	&-4.5	&-2.9\\\rowcolor{colororca2}
			PBE	&-22.8	&-43.4	&-76.9	&-94.1	&-116.2	&-122.5	&-63.0	&-17.5	&-30.9\\\rowcolor{colororca}
			PBE0	&-11.5	&-14.7	&-40.5	&-40.5	&-56.7	&-47.6	&19.3	&6.1	&13.7\\\rowcolor{colororca2}
			PW1PW	&-11.7	&-15.3	&-41.2	&-41.5	&-57.7	&-49.0	&17.8	&5.7	&12.8\\\rowcolor{colororca}
			PW6B95	&-13.2	&-15.2	&-41.1	&-39.2	&-54.8	&-44.2	&20.8	&6.2	&14.6\\\rowcolor{colororca2}
			PW91	&-23.3	&-44.3	&-78.3	&-96.0	&-118.5	&-125.4	&-66.3	&-18.3	&-32.7\\\rowcolor{colororca}
			PWP	&-24.7	&-47.5	&-83.3	&-103.0	&-127.5	&-136.5	&-79.7	&-20.9	&-38.6\\\rowcolor{colororca2}
			PWP1	&-13.1	&-18.5	&-46.2	&-48.5	&-66.8	&-60.2	&4.3	&3.1	&6.8\\\rowcolor{colororca}
			REVPBE	&-16.2	&-31.1	&-57.4	&-68.4	&-82.2	&-82.5	&-14.4	&-6.7	&-8.3\\\rowcolor{colororca2}
			RI-B2PLYP	&-11.4	&-12.6	&-38.8	&-36.3	&-53.1	&-42.7	&21.7	&11.3	&21.0\\\rowcolor{colororca}
			RI-PWPB95	&-13.2	&-12.2	&-39.7	&-35.2	&-53.4	&-40.1	&19.9	&9.7	&19.0\\\rowcolor{colororca2}
			RPBE	&-15.3	&-29.6	&-54.9	&-65.3	&-78.2	&-77.9	&-8.8	&-5.4	&-5.9\\\rowcolor{colororca}
			TPSS	&-17.2	&-31.7	&-59.0	&-69.6	&-85.2	&-84.7	&-22.4	&-7.5	&-10.5\\\rowcolor{colororca2}
			TPSS0	&-7.1	&-6.0	&-26.7	&-22.0	&-32.7	&-18.1	&51.2	&13.5	&28.8\\\rowcolor{colororca}
			TPSSh	&-13.0	&-21.0	&-45.6	&-49.8	&-63.5	&-57.3	&7.8	&1.4	&6.1\\\rowcolor{colororca2}
			X3LYP	&-10.9	&-15.0	&-38.4	&-38.7	&-50.9	&-42.7	&27.1	&8.1	&18.1\\\rowcolor{colororca}
			XLYP	&-16.6	&-32.9	&-58.6	&-70.8	&-82.9	&-84.9	&-16.7	&-6.2	&-8.1\\ \rowcolor{colornwchem}
			B1B95	&-13.7	&-15.3	&-41.5	&-39.3	&-55.5	&-44.6	&107.1	&5.9	&14.9\\\rowcolor{colornwchem2}
			BB1K	&-8.2	&-1.7	&-24.3	&-14.2	&-27.4	&-8.5	&172.6	&16.6	&34.7\\\rowcolor{colornwchem}
			BECKE97-3	&-11.7	&-11.6	&-35.5	&-31.8	&-44.8	&-32.7	&127.1	&8.7	&20.9\\\rowcolor{colornwchem2}
			BECKE97-D	&-15.5	&-29.8	&-55.8	&-19.4	&-80.2	&	&	&-3.2	&\\\rowcolor{colornwchem}
			FT97	&-14.5	&-27.7	&-50.1	&-10.2	&-68.4	&-67.3	&	&-4.0	&\\\rowcolor{colornwchem2}
			M05	&-9.2	&-8.4	&-34.9	&-30.6	&-48.8	&-34.5	&32.5	&10.8	&21.8\\\rowcolor{colornwchem}
			M05-2X	&-12.4	&-9.3	&-38.2	&-32.0	&-50.4	&-37.8	&30.3	&11.8	&20.4\\\rowcolor{colornwchem2}
			M06	&-12.2	&-11.2	&-38.7	&-34.7	&-54.3	&-42.2	&20.9	&8.3	&15.5\\\rowcolor{colornwchem}
			M06-2X	&-12.7	&-6.1	&-33.8	&-24.6	&-42.4	&-27.4	&36.3	&13.8	&25.7\\\rowcolor{colornwchem2}
			M06-HF	&-16.5	&-6.1	&-36.4	&-20.7	&-35.6	&-16.7	&52.1	&14.2	&25.3\\\rowcolor{colornwchem}
			M06-L	&-11.3	&-16.1	&-39.5	&-42.1	&-57.1	&-49.5	&	&2.7	&31.9\\\rowcolor{colornwchem2}
			MPW1B95	&-13.7	&-14.8	&-41.6	&-39.0	&-56.3	&-45.0	&110.9	&6.1	&14.4\\\rowcolor{colornwchem}
			MPW1B1K	&-8.5	&-1.9	&-25.1	&-15.0	&-29.1	&-10.3	&57.3	&16.2	&33.4\\\rowcolor{colornwchem2}
			PWB6K	&-7.1	&0.7	&-21.3	&-10.1	&-22.8	&-2.7	&66.0	&18.4	&37.2\\\rowcolor{colornwchem}
			VS98	&-11.5	&-21.5	&-45.9	&-54.0	&-70.8	&-68.6	&	&3.0	&25.9\\\rowcolor{colorg09}
			M05/Pople*	&-10.2	&-11.1	&-38.5	&-35.2	&-54.8	&-42.1	&24.4	&8.0	&17.0\\\rowcolor{colorg092}
			M06/Pople*	&-13.7	&-15.2	&-44.5	&-42.1	&-63.3	&-52.9	&11.0	&4.9	&8.2\\\rowcolor{colororca}
			PW6B95**	&-13.3	&-15.3	&-41.3	&-39.7	&-55.5	&-45.3	&19.4	&6.1	&13.9\\\rowcolor{colororca2}
			PBE0**	&-11.5	&-14.8	&-40.8	&-41.1	&-57.6	&-49.0	&17.6	&5.9	&\\\hline
		\end{tabular} 
		\caption{
			Software used for the calculations:  \color{orange}Orca 2.9.1\color{black},\color{cyan} NWChem 6.1.1\color{black}, \color{violet} G09 \color{black}
			Calculations were performed at functional/def2-QZVPP//B3LYP/def2-TZVPP level of theory
			“Pople*” means: Geometry optimization and single point energy calculation employing the 6-311+G(3d2f) Pople's basis set. “**” means: Geometry optimization and single point energy calculation employing the respective functional and the def2-QZVPP basis set.}
	\end{center}
\end{table*}

\cleardoublepage
\newpage

\section*{Elimination energy curves}
Elimination energy curves of fluorine or chlorine atom in kcal$\cdot$mol$^{-1}$. Blue curves represent the elimination of a fluorine atom from the molecule. Yellow curves represent triplet states for closed shell species and elimination of chlorine atom for open shell cases.
\vspace{0.5cm}

\includegraphics[scale=0.85]{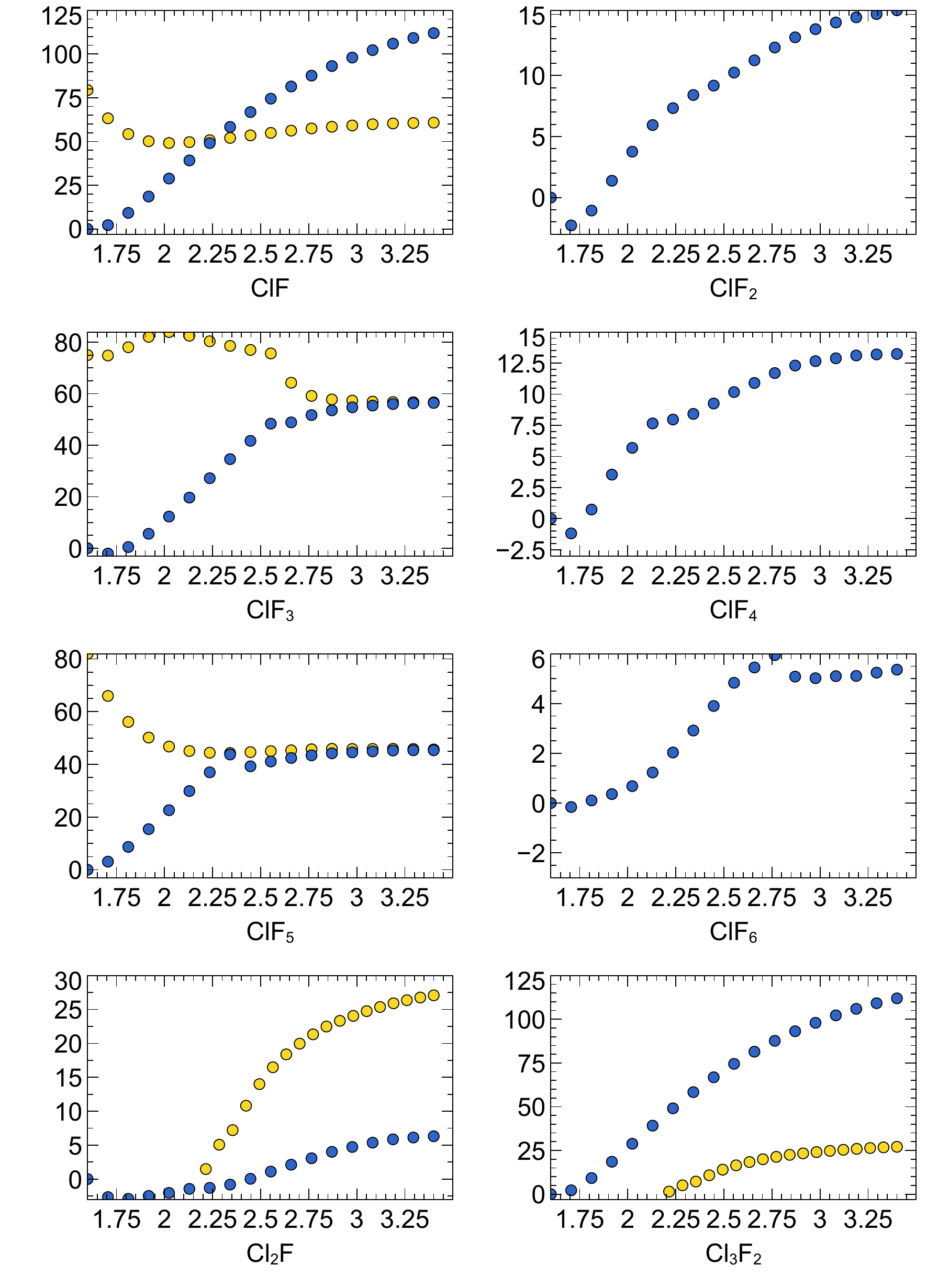}

\cleardoublepage
\newpage

\section*{Molecular geometries}

Cartesian coordinates of equilibrium geometries in Angstroms.
\vspace{0.5cm}

ClF
\begin{lstlisting}
F          0.00000        0.00000       -1.07420
Cl         0.00000        0.00000        0.56870
\end{lstlisting}
\vspace{0.3cm}

ClF$_2$
\begin{lstlisting}
Cl         0.00000        0.19203        0.00000
F          1.68993       -0.18140        0.00000
F         -1.68993       -0.18132        0.00000
\end{lstlisting}
\vspace{0.3cm}

ClF$_3$
\begin{lstlisting}
Cl        -0.00015       -0.35679       -0.00003
F         -1.72073       -0.29302        0.00003
F          1.72047       -0.29352        0.00003
F          0.00053        1.26048       -0.00000
\end{lstlisting}
\vspace{0.3cm}

ClF$_4$
\begin{lstlisting}
Cl        -0.00000       -0.00001        0.16609
F         -1.69202       -0.00306       -0.07865
F          1.69207        0.00302       -0.07865
F          0.00301       -1.69198       -0.07820
F         -0.00306        1.69204       -0.07823
\end{lstlisting}
\vspace{0.3cm}

ClF$_5$
\begin{lstlisting}
Cl        -0.00000       -0.00006       -0.29868
F          0.00007        0.00013        1.31348
F         -1.68176       -0.08541       -0.18734
F          1.68166        0.08522       -0.18740
F         -0.08530        1.68176       -0.18734
F          0.08533       -1.68159       -0.18723
\end{lstlisting}
\vspace{0.3cm}

ClF$_6$
\begin{lstlisting}
Cl         0.00001       -0.00001        0.00001
F         -1.49275       -0.22284        0.77860
F         -0.47344        1.56384       -0.46088
F          0.65726        0.62206        1.43657
F         -0.65628       -0.62182       -1.43621
F          1.49143        0.22231       -0.77824
F          0.47378       -1.56353        0.46014
\end{lstlisting}
\vspace{0.3cm}

ClF$_7$
\begin{lstlisting}
Cl        -0.00004       -0.00001        0.00000
F         -0.00006       -0.00001        1.58639
F          0.87003        1.53652        0.00056
F          1.73051       -0.35265       -0.00046
F         -1.19238        1.30245       -0.00045
F          0.19928       -1.75453        0.00018
F         -0.00005       -0.00001       -1.58638
F         -1.60726       -0.73175        0.00015
\end{lstlisting}
\vspace{0.3cm}

\newpage

Cl$_2$F
\begin{lstlisting}
F          1.33034        1.72925        0.00000
Cl         0.00000        0.55354        0.00000
Cl        -0.70430       -1.46903        0.00000
\end{lstlisting}
\vspace{0.3cm}

Cl$_3$F$_2$
\begin{lstlisting}
Cl        -0.00035        0.27530        0.00000
Cl        -2.15113       -0.24743        0.00000
Cl         2.15185       -0.24664        0.00000
F         -0.00035        0.20663        1.74339
F         -0.00035        0.20663       -1.74339
\end{lstlisting}

\end{document}